# Topological holography and storage with optical knots and links

*Ling-Jun Kong*, Jingfeng Zhang*, Furong Zhang, and Xiangdong Zhang$^§$*

Key Laboratory of advanced optoelectronic quantum architecture and measurements of Ministry of Education, Beijing Key Laboratory of Nanophotonics & Ultrafine Optoelectronic Systems, School of Physics, Beijing Institute of Technology, 100081 Beijing, China.

*These authors contributed equally to this work. $^§$*Author to whom any correspondence should be addressed. E-mail: zhangxd@bit.edu.cn; konglj@bit.edu.cn
.

## Abstract

**After more than 70 years of development, holography has become an essential tool of modern optics of many applications. In fact, for various applications of different kinds of holographic techniques, stability and anti-jamming ability are very important. Here, we introduce optical topological structures into holographic technology and demonstrate an entirely new concept of optical topological holography to solve stability and anti-jamming problems. Based on the optical knots and links, the topological holography is not only developed in theory, but also demonstrated in experiment. In addition, we establish a new topological holographic coding by regarding each knotted/linked topological structure as an information carrier. Due to the variety of knotted and linked structures and their characteristics of topological protection, such coding can have high capacity as well as robust properties. Furthermore, with writing the hologram into the liquid crystal, robust information storage of three-dimensional topological holography has been realized.**

## 1. Introduction

The concept of holography was introduced by Gabor in 1948[1]. In 1960s, Kozma and Kelly combined computer and spatial frequency domain filtering technology to design matched spatial filters artificially[2], which laid the foundation for computer-generated holography. In computer-generated holography, the recording process is simulated by computers and the reconstruction can be realized by loading the computer-generated holograms on the display devices (such as liquid crystal display) with coherent light illumination.

With the development of computer and optoelectronic technology, holography has been an essential tool of modern optics of many applications for three-dimensional displays[3–9], microscopy[10,11], optical encryption[12,13], information storage[14] and so on. To date, the different degrees of freedoms (DOFs) of light, including polarization[15–17], wavelength[5,18–21], time[22,23], and orbital angular momentum (OAM)[24,25], have been utilized to carry independent information channels for high-capacity holographic systems. In fact, for various applications of different kinds of holographic techniques, stability and anti-jamming ability are very important. However, at present, there is no good way to realize the stability and anti-jamming ability in optical holography systems.

On the other hand, in recent years the study of topological structures has attracted great attention because these light fields are robust against perturbations[26–35]. Knots and links are typical topological structures, and they are defined mathematically as closed curves in three-dimensional (3D) space associated with physical strings[36,37]. At present, knots and links has been observed in plasma[38], quantum and classical fluids[38], quantum and classical field theory[39–41], liquid crystal[42,43], sound wave[44], and optical fields[29–31,34,45–48]. In topology, two knots/links are equivalent if one can be transformed into the other via the continuous deformation without cutting the lines or permitting the lines to pass through itself. In other words, even if a knotted/linked structure may be deformed and distorted due to the external disturbances, its topological invariants remain unchanged, which fully exhibits the robust topological properties. The question is whether the topological structures, such as knots and links, can be introduced into holography to solve the stability and anti-jamming ability problems against external perturbations.

In this work, we demonstrate an entirely new concept of optical topological holography by introducing topological structures into the holographic technology. Based on the optical knots and links, we build the theory of the topological holography and realize the optical topological holography in experiment. By using optical knotted/linked structure as an information carrier, we propose and experimentally demonstrate a topological holographic coding. Because there are many kinds of knots and links, and each of them is topologically protected, such 3D coding can have not only high capacity, but also robust properties. By using liquid crystal materials to record the information carried by the 3D knotted and linked structures, the topological holographic storage is also realized in experiments.

## 2. Topological holography based on the optical knots and links

To achieve topological holography, one should not only be able to generate each knot or link very well, but also need to generate an array of 3D knotted/linked structures. In the array, various knotted/linked structures should be separated from each other in space without overlap, which ensures the topological characteristics of each knot and link to be well preserved. At present, there are two methods to generate an optical knot or link. One is based on singularity optics[31,45,48,49], the other is based on shaping the light beams along curves in three dimensions[50]. In the singularity optics, when the knotted and linked structures are complicated, the distance between adjacent singularities may become too small for the detector to distinguish them, so that it is very difficult for them to be reconstructed experimentally. Here we generate 3D knotted/linked structures with the later method.

Any knotted/linked structure can be described by a 3D curve equation $\mathbf{K}(t) = (x(t), y(t), z(t))$, where $(x, y, z)$ is the coordinate in real 3D space with $t \in [0, t_{max}]$. For example, when $x(t) = -\cos(t) + 2.5\cos(2t)$, $y(t) = 1.5\sin(3t)$, $z(t) = \sin(t) + 2.5\sin(2t)$ and $t_{max} = 2\pi$, $\mathbf{K}(t)$ depicts the trefoil knot and is marked as $3_1$ knot with Alexander-Briggs notation[36]. According to the method of shaping the light beams along curves in three dimensions as described in Ref.[50], the knotted and linked structures with the relative maximum points of the amplitude distribution can be obtained theoretically. The detailed calculation process is summarized in Supporting Information Note 1. In experiments, we can use the spatial light modulator (SLM) to show such a phenomenon. That is, we encode the phase distribution on the SLM, measure the intensity distribution of the light field at different transverse planes by illuminating the SLM with the fundamental mode Gaussian beam. Then, the knotted and linked topological structures can be reconstructed. More details on the experimental setup is provided in the Experimental Section. In Supporting Information Note 2, we provide calculated and experimental results of reconstructed optical topological structures for various knots and links. The topological invariants of each knot or link, the linking number and Alexander polynomial, are also calculated and provided. All experimental results are in good agreement with the theoretical results.

Now, we discuss how to use the 3D knot or link to construct an array for realizing topological holography. Here, we introduce parallel moving grating (PMG) into the phase distribution of complex amplitude (PCA), $\arg[C_A(\varepsilon, \eta)]$, to control the position of the 3D knotted/linked structure accurately, as shown in **Figure** 1a. The function of PMG can be described as $e^{-j\frac{2\pi}{\lambda z_0}(\alpha_i \varepsilon + \beta_j \eta)}$. Diffractive factor $(\alpha_i, \beta_j)$ determines the center position $(x_i, y_j, 0)$ of the 3D knotted/linked structure in the $i$-th row and the $j$-th column of the array. Based on the PCA and PMG, we construct the hologram $H_{ij}(\varepsilon, \eta) = e^{-j \cdot \arg[C_A(\varepsilon, \eta)]} e^{-j\frac{2\pi}{\lambda z_0}(\alpha_i \varepsilon + \beta_j \eta)}$ to generate a knotted/linked structure at a specified position $(i, j)$ of the array. Then, the hologram of topological holography can be constructed by the trivial superposition $H_{ij}(\varepsilon, \eta)$ as

$$\mathcal{H}_{MN}^{TH}(\varepsilon, \eta) = \sum_{j=1}^{N} \sum_{i=1}^{M} H_{ij}(\varepsilon, \eta) = \sum_{j=1}^{N} \sum_{i=1}^{M} e^{-j \cdot \arg[C_A(\varepsilon, \eta)]} e^{-j\frac{2\pi}{\lambda z_0}(\alpha_i \varepsilon + \beta_j \eta)}. \tag{1}$$

Here, *M* and *N* represent the total number of rows and columns of the array, respectively. Then, the light field containing the knotted and/or linked structures can be calculated by

$$E_{MN}^{TH}(x,y,z) = \frac{e^{j2\pi(z+z_0)/\lambda}}{j\lambda(z+z_0)} \int\int_{-\infty}^{\infty} E_{in}(\varepsilon,\eta)\mathcal{H}_{MN}^{TH}(\varepsilon,\eta)e^{j\frac{\pi}{\lambda(z+z_0)}[(\varepsilon-x)^2+(\eta-y)^2]}d\varepsilon d\eta.$$

$$= \frac{e^{j2\pi(z+z_0)/\lambda}}{j\lambda(z+z_0)}[E_{in}(x,y)\mathcal{H}_{MN}^{TH}(x,y)]$$

$$\circledast\, e^{j\frac{\pi(x^2+y^2)}{\lambda(z+z_0)}}. \qquad (2)$$

Here, $z_0$ represents the distance between the hologram plane $(\varepsilon,\eta)$ and the plane with $z=0$ in the $(x,y,z)$ coordinate (as shown in Figure 1a). $E_{in}(\varepsilon,\eta)$ represents the incident light field. In our theoretical simulations and experiments, $E_{in}(\varepsilon,\eta)$ is set to be a fundamental Gaussian mode whose beam waist is large enough to cover the hologram, and the sign $\circledast$ denotes convolution. Now, we take the $3_1$ knot as an example to provide theoretical results. From Eq. (1) and Eq. (2), the calculated results of the intensity distributions at different transverse planes in $(x,y,z)$ coordinate are shown in Figure 1b. Here the numbers of rows and columns of the array, *M* and *N*, are set to be 6. The enlarged view of a unit at the plane with $z=0$ is shown in Figure 1c. It is seen that there are six extreme maximum points circled with dashed circles. The locations and the number of the bright points change along $z$ axis. By connecting the extreme maximum points on the different transverse planes, the array of isolated optical $3_1$ knotted lines can be obtained, which is shown Figure 1d. The structural array composed of $3_1$ knots clearly appears.

In experiments, we encode the phase distribution $\mathcal{H}_{MN}^{TH}$ into the SLM, illuminate the SLM with the fundamental mode Gaussian beam, and measure the intensity distributions of the light field at different transverse planes similar to the case of the single knot. The experimental results of intensity distributions are shown in Figure 1e. The corresponding enlarged view of a unit at the plane with $z=0$ and array of $3_1$ knotted lines are plotted in Figure 1f and Figure 1g, respectively. Comparing these experimental results with the theoretical results, it is found that the agreements between them are very well. If we take other 3D knotted/linked structures to perform experiments, similar phenomena can be observed. The calculated and experimental results for the array of $6_1^2$ linked lines are given in Supporting Information Note 3. These results confirm the feasibility of the holography with topological light fields.

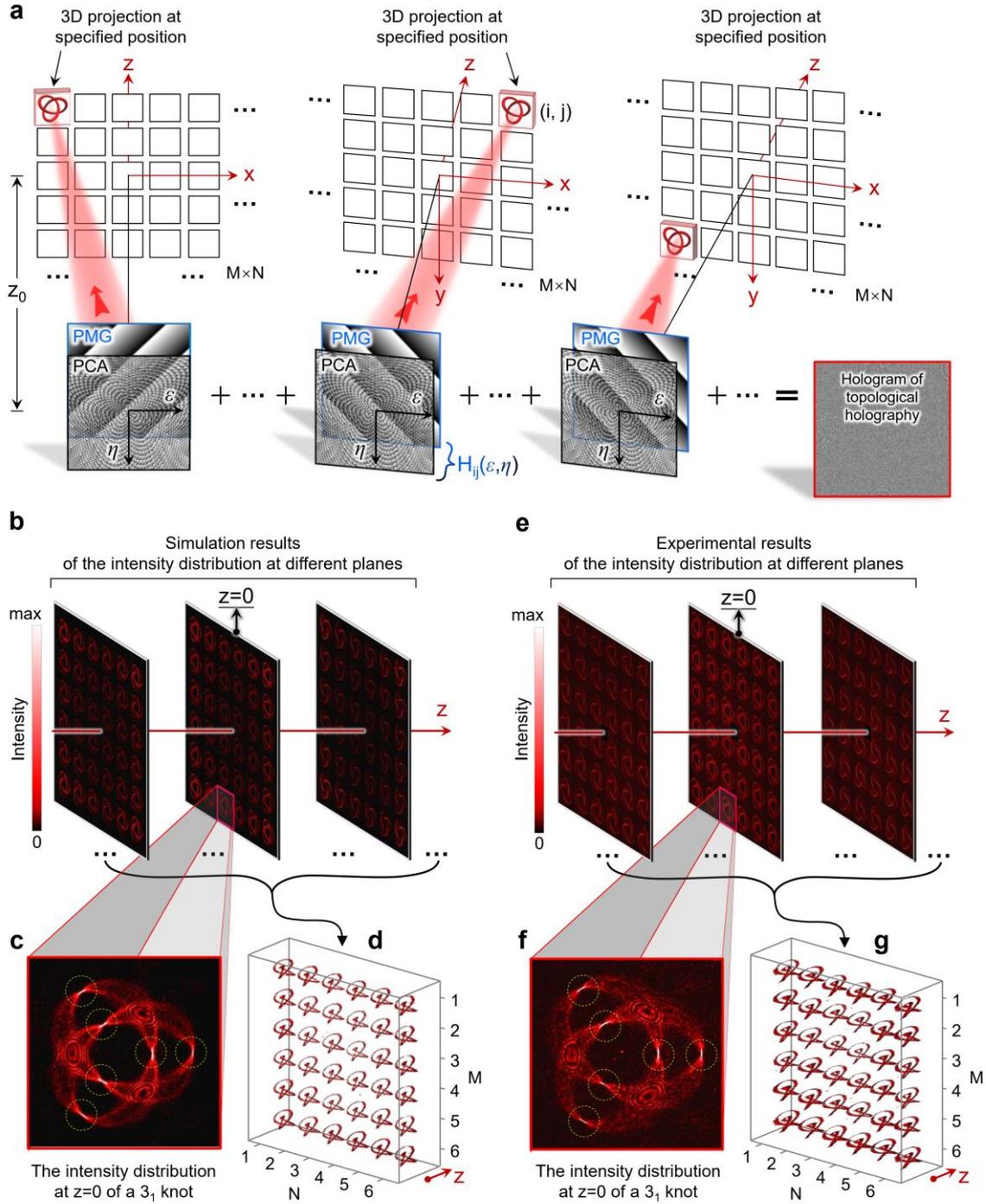

**Figure 1.** Principle, simulation results, and experimental results of optical topological holography. a) Construction of hologram for topological holography with *MN* multi-knot/link projections. PCA represents the phase distribution of complex amplitude used for generating 3D knotted/linked structure. PMG represents the parallel moving grating, which is described with $e^{-j2\pi(\alpha_i \varepsilon + \beta_j \eta)}$ and can move the center position of the 3D knotted/linked structure from $(0,0,0)$ to $(x_i, y_j, 0)$. b) Simulation results of the intensity distributions at different transverse planes in $(x, y, z)$ coordinate with $M = 6$, $N = 6$. c) The enlarged view of one part at $z = 0$ plane shown in (b). d) Simulation results of optical topological holography. e)-g) Corresponding experimental results of (b)-(d), respectively.

In the above topological holography, the array is composed of one certain kind of topological structure. That is, one kind of PCAs is used to construct the hologram. In fact, we can use various PCAs, corresponding to different knotted/linked structures, to construct the hologram. In such a case, different topological structures can be projected into different positions of the array. Thus, the topological holography composed of different topological structures can also be realized. One example is shown in **Figure** 2, where $0_1$, $3_1$, $5_1$, $0_1^2$, $2_1^2$, $0_1^3$, $6_1^2$, and $6_3^3$ knots/links are used. Figure 2a corresponds to the theoretical results and Figure 2b to the experimental results. Both theoretical and experimental results show the good effect of holography with different 3D knotted/linked structures. The topological invariants (the linking number and Alexander polynomial) of each corresponding topological structure in the theoretical and experimental results are consistent with each other. It is very interesting that the topological holography composed of various topological structures can be used to establish a novel topological holographic coding scheme, because each knotted/linked structure can be regarded as a carrier of information.

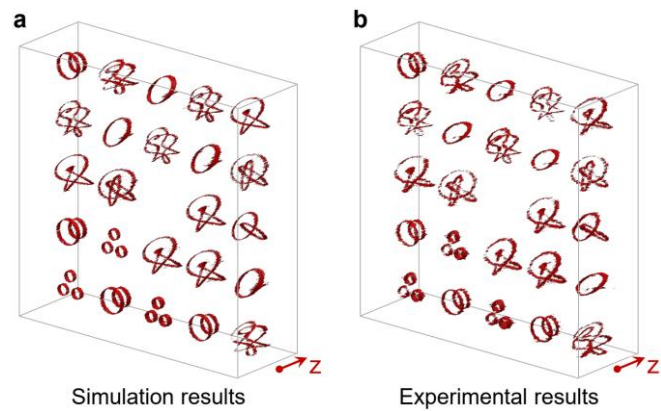

**Figure 2.** Optical topological holography with various knotted and linked structures. a) and b), Simulation and experimental results of topological holography with knots and links of $0_1$, $3_1$, $0_1^2$, $2_1^2$, $0_1^3$, $5_1$, $6_1^2$, and $6_3^3$.

## 3. Coding based on the topological holography

To prove the topological holographic coding scheme, we select ten kinds of knots and links to encode the ten digits of Arabic numerals 0 to 9, as shown in **Figure** 3a. Then, we use two digits of the Arabic numerals from 01 to 26 to represent alphabet letters from A to Z, respectively. In order to make the description clearer, an example is taken to demonstrate the coding scheme based on the topological holography (Figure 3b). The target message is the question 'Where is my key', which can be transformed into the number combinations '23 08 05 18 05 09 19 13 25 11 05 25'. According to the knots and links code chart shown in Figure 3a, the above number combinations can be transformed into an array of knots and links as '$0_1^2$, $2_1^2$, $0_1$, $6_1^2$, $0_1$, $0_1^3$, $3_1$, $6_1^2$, $0_1$, $0_1^3$, $0_1$, $6_3^3$, $3_1$, $6_3^3$, $3_1$, $2_1^2$, $0_1^2$, $0_1^3$, $3_1$, $3_1$, $0_1$, $0_1^3$, $0_1^2$, $0_1^3$'. Based on the process shown in Figure 1a, Alice can construct the hologram of topological holography for projecting these 24 knots and links. Then, Alice sends the desired hologram to Bob, the receiver of the information. Bob reconstructs the experimental results of the topological holography, calculates the topological invariants (the linking number and Alexander polynomial) of each topological structure, identifies the type of knots and links, transforms them into the alphabet letters, and obtains the target messages 'Where is my key'.

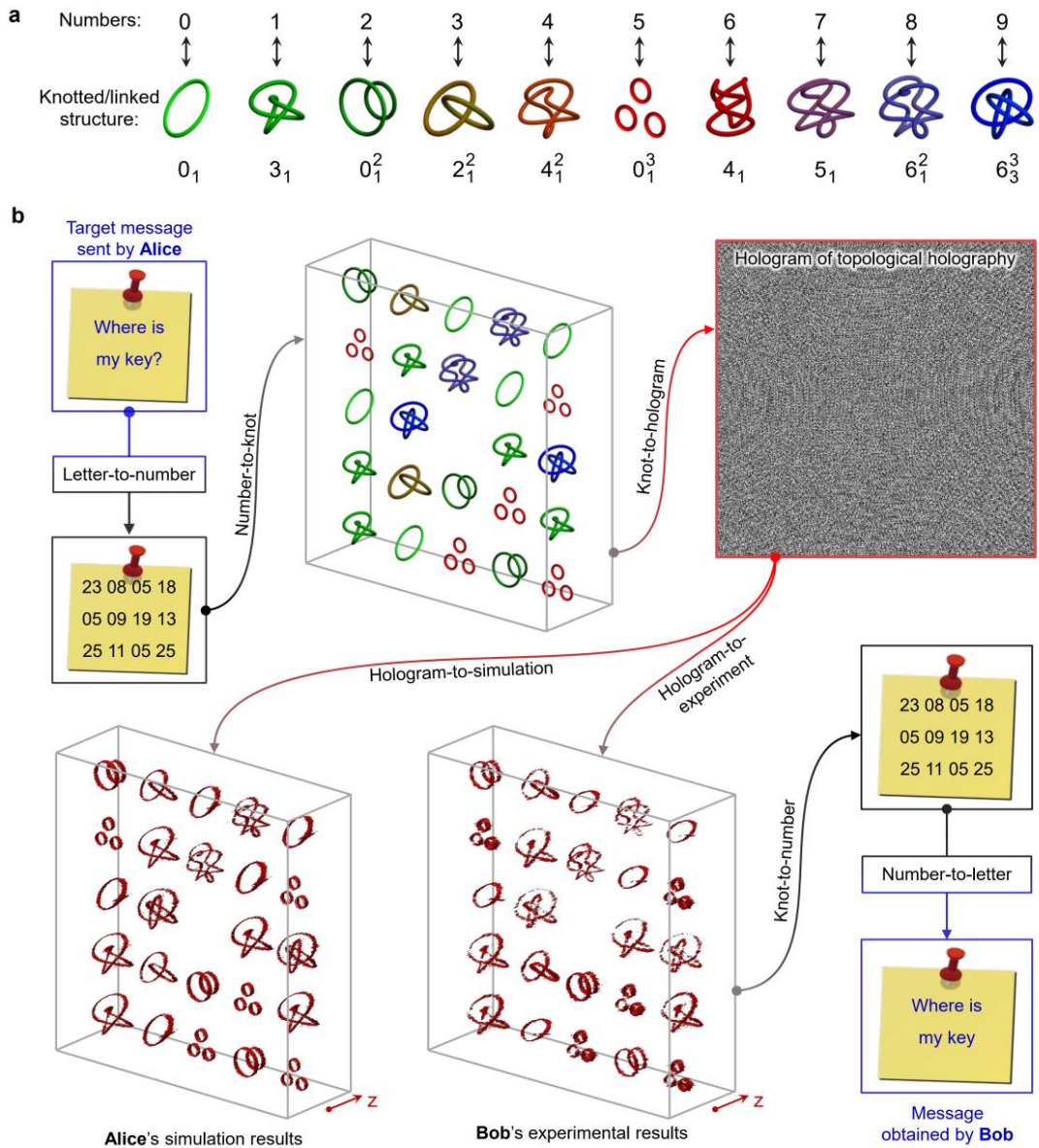

**Figure 3.** Experimental demonstration of coding based on the topological holography. a) The code chart for describing the relationship between knots/links and Arabic numerals. b) The process of information transmission in the coding based on the topological holography. Alice translates her target messages 'Where is my key' into the number combinations '23 08 05 18 05 09 19 13 25 11 05 25', which will be translated into an array of knots and links as '$0_1^2$, $2_1^2$, $0_1$, $6_1^2$, $0_1$, $0_1^3$, $3_1$, $6_1^2$, $0_1$, $0_1^3$, $0_1$, $6_3^3$, $3_1$, $6_3^3$, $3_1$, $2_1^2$, $0_1^2$, $0_1^3$, $3_1$, $3_1$, $0_1$, $0_1^3$, $0_1^2$, $0_1^3$' based on the code chart shown in (a). Alice constructs the hologram of topological holography, and sends it to Bob. Bob reconstructs the experimental results of the topological holography, calculates the topological invariants (the linking number and Alexander polynomial) of each topological structure, identifies the 3D knotted/linked structures, translates them into number combinations, and obtains target messages 'Where is my key'.

## 4. Information storage of 3D topological holography

By storing the information contained in topological holography into a recording medium, the information storage of 3D topological holography can be realized. Here, we select the liquid crystal (LC) as the recording medium. The LC is a typical birefringent material and it can be used to introduce the space-variant Pancharatnam-Berry (PB) phase[51]. The relationship between the introduced PB phase ($\phi$) and the fast axis orientation angle ($\theta$) of LC is $\phi = 2\theta$. Therefore, the phase profiles of the hologram of topological holography could be achieved by controlling the fast axis orientation. The orientation control can be realized by laser direct writing[52]. For example, the target message 'It is in the bag' is encoded by the knots and links combinations of '$0_1$, $6_3^3$, $0_1^2$, $0_1$, $0_1$, $6_3^3$, $3_1$, $6_3^3$, $0_1$, $6_3^3$, $3_1$, $4_1^2$, $0_1^2$, $0_1$, $0_1$, $6_1^2$, $0_1$, $0_1^3$, $0_1$, $0_1^2$, $0_1$, $3_1$, $0_1$, $5_1$'. Then, the corresponding hologram of topological holography can be constructed and written into a LC sample (shown in **Figure** 4a). The work area of the sample contains 1000×1000 pixels with the pixel pitch of 10μm. By measuring the intensity distribution of diffracted light field behind the LC, the experimental results are reconstructed and shown in Figure 4b. After calculating the topological invariants (the linking number and Alexander polynomial) of each topological structure, identifying the type of knots and links, and transforming them into the alphabet letters, the information 'It is in the bag' stored in the sample is extracted (Figure 4c).

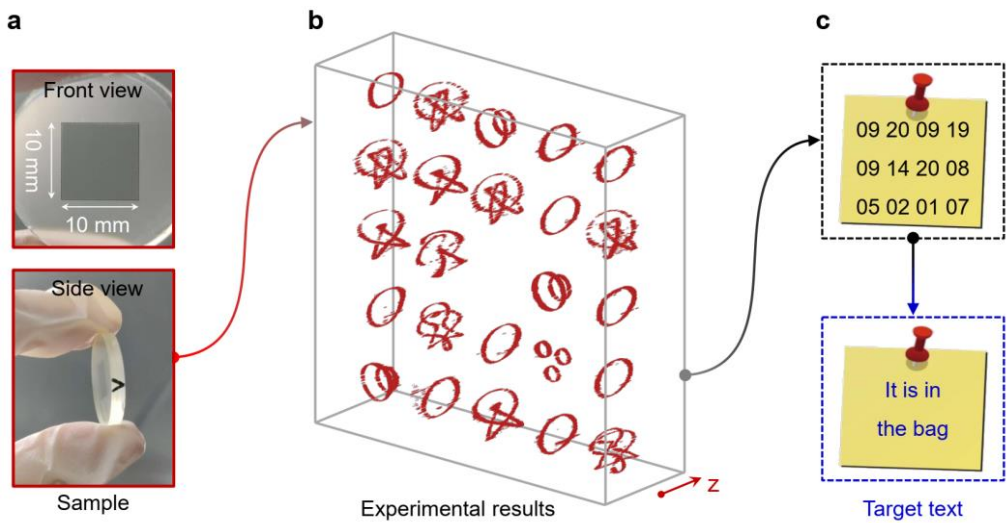

**Figure 4.** Experimental demonstration of optical storage of topological holography. a) Pictures of the LC sample. The work area is framed by a dotted square. b) Reconstructed topological 3D knotted/linked structures with the sample. c) The 3D knotted/linked structures are transformed into number combinations '09 20 09 19 09 14 20 08 05 02 01 07' based on the knots and links code chart shown in Figure 3a, and translated into the messages 'It is in the bag'.

## 5. Discussion

We have realized the topological holography by introducing the 3D optical knotted and linked topological structures into the holographic technology. Not only the theory of optical topological holography has been developed, but also the experimental demonstration has been given. With regarding the knotted/linked topological structure as an information carrier, we have established a novel topological holographic coding scheme. On the other hand, because the 3D optical knotted/linked structures correspond to the concept of topology in mathematics, and they contain topological invariants, such as linking number, Alexander polynomial, Jones polynomial, genus number, etc. These topological invariants are preserved under continuous deformations. Therefore, in optical topological holographic coding, 3D optical knotted/linked structures have robustness to certain deformation or noise. We deep study the robustness of 3D optical knotted/linked structures for the random phase and the deformation of hologram. Our results show that the topological structure can keep unchanged with the random phase in the range of [0,15rad] and some deformations of holograms (more discussions can be found in Supporting Information Note 4). That is to say, our topological holography scheme has a good robustness and can solve the stability and anti-interference problems of holographic technology.

It should be noted that our coding scheme based on our topological holography is different from previous coding method in Refs.[48, 49], where only one knot or link has been used to carry information. In our topological holographic coding, the knot and link arrays have been utilized for encoding information and the capacity of information can increase as the number of the knot and link increases. Compared with other holographic technologies, the introduce of the 3D topological structures makes our method have some unique advantages. In our method, the 3D knots and links can be regarded as a novel "DOF". Different knots and links correspond to different "dimensions". Since there are many types of knots and links, this novel "DOF" has a high "dimensionality", which makes the topological holographic coding scheme have a high capacity.

Although the above results have been obtained in optical systems, our method can be extended to sound, x-ray, electron system and so on. Furthermore, by using the LC as the recording medium, we have realized the 3D topological holographic information storage. In recent years, metasurface, as an integrated platform, has greatly expanded the ability of generation of optical knotted and linked topological structures[53]. Therefore, it is also feasible to realize topological holographic coding and storage based on metasurfaces. At present, the reconstruction of 3D knotted and linked structure has been accomplished by detecting a large number of light intensity distribution in the three-dimensional space, which is the main limitation of our technology application. The development of machine learning may provide help to overcome this limitation and empower our topological holographic technique to photonic applications in the next decade[54].

## 6. Experimental Section.

Our experimental setup is shown in Figure S3 in Supporting Information Note 2. The input light is a collimated fundamental mode Gaussian beam with the wavelength of 633 nm beam. The waist of the input beam is large enough to cover the entire working area of spatial light modulator (SLM). The SLM is a phase-only reflective modulator (Holoeye Pluto-2-NIR-015) with 1920×1080 pixels and a pixel pitch of 8μm. To display the relevant parameters more clearly, we show it in the form of transmissive SLM, which will not affect the demonstration of physical principles. The detector is a camera beam profile (Thorlabs, BC106N-VIS/M).

**Supporting Information**

Supporting Information is available from the Wiley Online Library or from the author.

**Acknowledgements**

This work was supported by the National key R & D Program of China under Grant No. 2017YFA0303800, National Natural Science Foundation of China (12004038, 11904022, and 11974046).

**Conflict of Interests**

The authors declare that they have no conflict of interest.

**Data Availability Statement**

The data that support the findings of this study are available from the corresponding author upon reasonable request.

# Supporting Information

# Laguerre Gaussian mode holography for ultra-secure encryption

Furong Zhang[*], Ling-Jun Kong[*], Zhuo Zhang, Jingfeng Zhang, and Xiangdong Zhang[$]

*Key Laboratory of advanced optoelectronic quantum architecture and measurements of Ministry of Education, Beijing Key Laboratory of Nanophotonics & Ultrafine Optoelectronic Systems, School of Physics, Beijing Institute of Technology, 100081 Beijing, China.*

*[*]These authors contributed equally to this work. [$]*Author to whom any correspondence should be addressed. E-mail: zhangxd@bit.edu.cn; konglj@bit.edu.cn*

**Supporting Information Note 1. The calculation method and results for generation of optical knots and links with the relative maximum points of the amplitude distribution.**

In this section, we describe the generation of optical knots and links with the relative maximum points of the amplitude distribution. Knots and links are defined mathematically as closed curves in three-dimensional space. Therefore, we can mathematically describe a knot/link with a 3D curve equation $\mathbf{K}(t) = (x(t), y(t), z(t))$. Here $x$, $y$, and $z$ are the coordinates in real 3D space. The value range of the variable $t$ is $[0, t_{max}]$. Based on the theory of shaping the light beams along curves in three dimensions [1], a light beam, whose intensity distribution follows the prescribed 3D knotted/linked curve $\mathbf{K}(t)$ in the $(x, y, z)$ space, can be designed. The complex amplitude of this light beam, $C_A(\varepsilon, \eta)$, can be calculated as

$$C_A(\varepsilon, \eta) = \exp\left[-\frac{j\pi}{\lambda z_0}(\varepsilon^2 + \eta^2)\right] \int_0^T |\mathbf{K}'(t)| \Phi(\varepsilon, \eta, \mathbf{K}(t)) \varphi(\varepsilon, \eta, \mathbf{K}(t)) dt, \qquad (S1)$$

where

$$|\mathbf{K}'(t)| = \sqrt{x'(t)^2 + y'(t)^2 + z'(t)^2}, \qquad (S1a)$$

$$\Phi(\varepsilon, \eta, \mathbf{K}(t)) = \exp\left\{\frac{j}{\omega_0^2}[\eta x(t) - \varepsilon y(t)] + \frac{j}{\omega_0^2}\int_0^T \left[x(\tau)\frac{dy(\tau)}{d\tau} - y(\tau)\frac{dx(\tau)}{d\tau}\right]d\tau\right\}, \qquad (S2b)$$

$$\varphi(\varepsilon, \eta, \mathbf{K}(t)) = \exp\left\{-\frac{j\pi}{4\lambda z_0^2}[\varepsilon - x(t)]^2 + [\eta - y(t)]^2 z(t)\right\}. \qquad (S2c)$$

Here, $\omega_0$ is a constant and $\lambda$ represents the wavelength, $(\varepsilon, \eta)$ are the spatial coordinates at the hologram plane, $z_0$ is the distance between $(\varepsilon, \eta)$ and the center of the 3D knotted and linked structure. The term $e^{-j\frac{\pi}{\lambda z_0}(\varepsilon^2+\eta^2)}$ plays the function of a Fresnel zone plate with the focal length $f = z_0$ and controls the position of the 3D knotted and linked structure. The term $|\mathbf{K}'(t)|$ guarantees a uniformly distributed intensity along the knotted curve. The term $\Phi(\varepsilon, \eta, \mathbf{K}(t))$ controls the position of each focused spot in the focal plane. While the term $\varphi(\varepsilon, \eta, \mathbf{K}(t))$ controls the focusing distance through a quadratic phase function. According to diffraction theory, the light field that contains the 3D knotted topological structure can be calculated by the following equation:

$$F_K(x,y,z) = \frac{e^{j2\pi(z+z_0)/\lambda}}{j\lambda(z+z_0)} \int\int_{-\infty}^{\infty} C_A(\varepsilon,\eta) e^{j\frac{\pi}{\lambda(z+z_0)}[(\varepsilon-x)^2+(\eta-y)^2]} d\varepsilon d\eta$$

$$= \frac{e^{j2\pi(z+z_0)/\lambda}}{j\lambda(z+z_0)} e^{j\frac{\pi}{\lambda(z+z_0)}(x^2+y^2)} \int\int_{-\infty}^{\infty} \left[ C_A(\varepsilon,\eta) e^{j\frac{\pi}{\lambda(z+z_0)}(\varepsilon^2+\eta^2)} \right] e^{-j\frac{2\pi}{\lambda(z+z_0)}(x\varepsilon+y\eta)} d\varepsilon d\eta. \quad (S3)$$

This equation shows that the light field contained 3D knotted/linked structure at the plane of $z+z_0$ is the Fourier transform of the hologram $C_A(\varepsilon,\eta)e^{j\frac{\pi}{\lambda(z+z_0)}(\varepsilon^2+\eta^2)}$.

We take the $3_1$ knot as an example. When $x(t) = -\cos(t) + 2.5\cos(2t)$, $y(t) = 1.5\sin(3t)$, and $z(t) = \sin(t) + 2.5\sin(2t)$ with $t \in [0, 2\pi]$, the $\mathbf{K}(t)$ represents the $3_1$ knot. Figure S1a shows the 3D curve of $3_1$ knot. The complex amplitude used for generating the prescribed 3D $3_1$ knotted/linked curve in the $(x,y,z)$ space can be calculated with Eq. (S1). The amplitude and phase distributions of the complex amplitude are shown in Figure S1b and 1c, respectively. The light field that contains the 3D $3_1$ knotted topological structure can be calculated with Eq. (S3). Simulation results of the intensity distributions at different transverse planes in $(x,y,z)$ coordinate as shown in Figure S1d. The extreme maximum points are circled with dashed circles. The locations and the number of the extreme maximum points evolve along $z$ axis. By connecting the extreme maximum points on the different planes, the isolated optical $3_1$ knotted line can be reconstructed (Figure S1e).

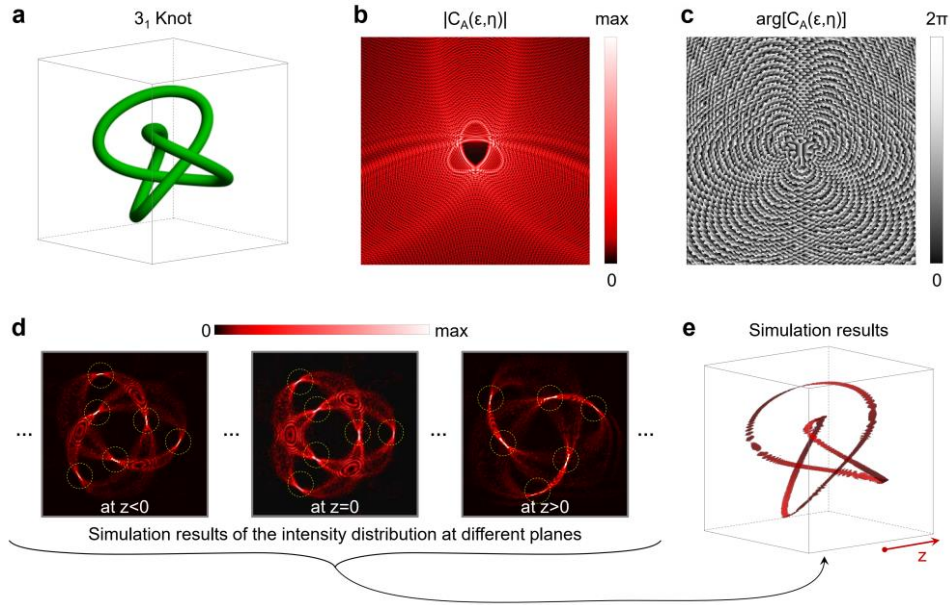

**Figure S1. Calculation of the optical topological structures. a**, Three-dimensional curve of $3_1$ knot. **b**(**c**), The intensity (phase) distribution calculated with Eq. (S1). **d**, Simulation results of intensity distributions calculated with the Eq. (S3) for $3_1$ knot at different transverse planes. **e**, Simulation result of reconstructed optical topological structure of $3_1$ knot.

**Supporting Information Note 2. The experimental results and setup for generating the optical knots and links**

In experiment, for maximum generality and best results, one should encode the phase distribution of $C_A(\varepsilon,\eta)$ ($\arg[C_A(\varepsilon,\eta)]$) on the SLM and input the light field with the amplitude distribution of $|C_A(\varepsilon,\eta)|$. However, because the amplitude distribution is too complex (as shown in Figure S1b), it is not easy to be generated in experiment. Luckily, our studies show that the generation of optical knotted and linked structures has a very good robustness to amplitude distribution of incident field. Here, we still take the $3_1$ knot as an example. For the first condition, we use a uniform flat top beam (shown in Figure S2a1) as the input light beam, the calculated results with Eq. (S3) demonstrate that the isolated optical $3_1$ knotted structure can still be reconstructed (Figure S2b1). The corresponding intensity distributions at the plane of $z = 0$ is shown in Figure S2c1. For the second condition, the input amplitude distribution is set to be a fundamental mode Gaussian beam (Figure S2a2), the isolated optical $3_1$ knotted structure can still be reconstructed (Figure S2b2). Figure S2c2 represents the intensity distributions at the plane of $z = 0$. By comparing Figure S1e, 2b1, and 2b2, one can find that the reconstructed optical knot structures are almost the same in these three cases, which implies that the generation of optical knot structure has a good robustness to the amplitude distribution of incident field. Therefore, in our experiment, to reduce the difficulty in experiment, the fundamental mode of Gaussian beam is used as the input light beam to produce the topological knotted/linked structures.

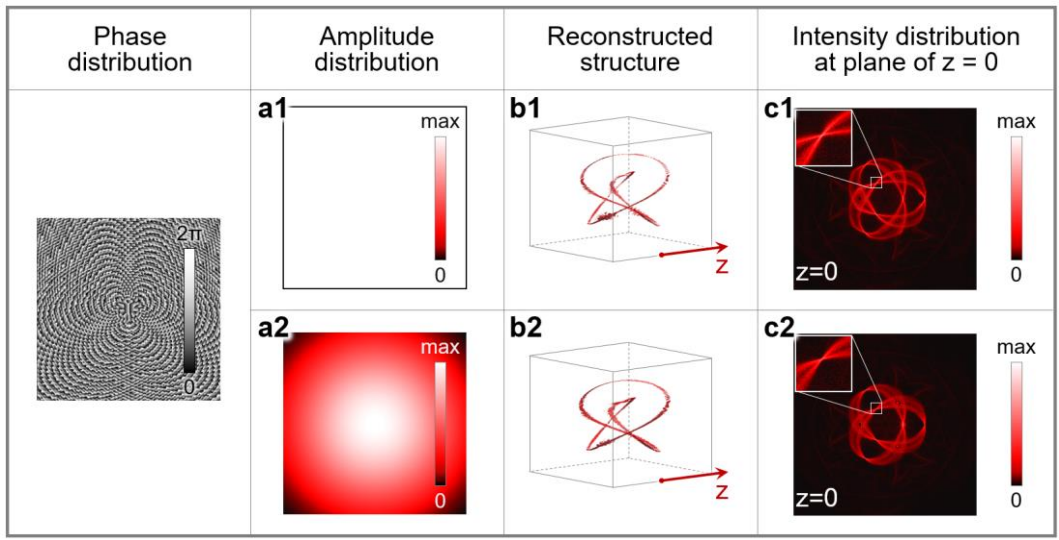

**Figure S2. Effect of amplitude distribution on the generation of optical knots and links. a1** and **a2** are input intensity distributions, uniform flat top beam, and fundamental mode Gaussian beam, respectively. **b1** and **b**3, Reconstructed topological structures of optical knot $3_1$ based on the relative maximum points of the amplitude distribution. **c1** and **c2**, Intensity distribution at plane of $z = 0$.

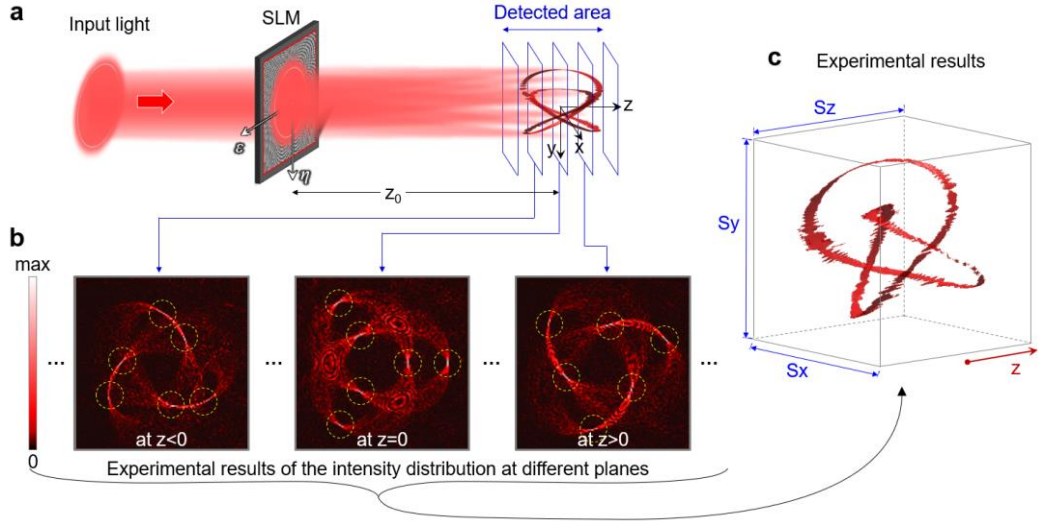

**Figure S3. Generation of the optical topological structures. a**, Experimental setup for generation of the optical topological structures. **b**, Measured intensity distributions for knot $3_1$ at different transverse planes. The extreme maximum points are circled with dashed circles. **c**, Reconstructed optical topological structure of knot $3_1$ with the measured intensity distributions at different transverse planes.

Our experimental setup for generating the optical knots and links is shown in Figure S3a. The input light is a collimated fundamental Gaussian beam with a beam waist large enough to cover the entire working area of spatial light modulator (SLM). The SLM is used for encoding the phase hologram. The plane of SLM is defined as hologram plane $(\varepsilon, \eta)$. The space of the knotted or linked structures is described with the $(x, y, z)$ coordinate system. The center of the knotted or linked structure is near the $z = 0$ plane. $z_0$ represents the distance between the hologram plane $(\varepsilon, \eta)$ and $z = 0$ plane. We encode the phase distribution of $C_A(\varepsilon, \eta)$ into the SLM, measure the intensity profile of the light field at different transverse planes in $(x, y, z)$ coordinate. The experimental results are shown in Figure S3b. By connecting the extremely bright points on the different planes, the isolated optical $3_1$ knotted lines can be obtained, which corresponds to the experimental results shown in Figure S3c. Results for more knots and links are shown in Figures S4 and S5. The topological invariants of each knot or link, the linking number and Alexander polynomial, are also calculated and provided.

To avoid crosstalk of these structures, one need to properly set the size of the knotted or linked structure ($S_x$, $S_y$, and $S_z$, as shown in Figure S3c) and the distance between two adjacent structures ($D$). When the crosstalk among the different part of the same structure is too large, especially for complex structures, we can increase the parameters, $S_x$, $S_y$, and $S_z$. When the crosstalk between two adjacent structures is too large, we can increase the $D$. In our experiment, $S_x \approx 5mm$, $S_y \approx 5mm$, $S_z \approx 120mm$, and $D \approx 4mm$.

| Knot/Link: | arg[$C_A(\varepsilon,\eta)$] PCA | Simulation results | Experimental results | Linking number | Alexander polynomials |
|---|---|---|---|---|---|
| $0_1$ 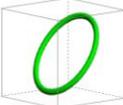 | 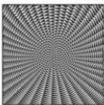 | 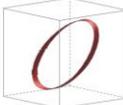 | 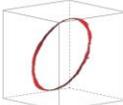 | - | 1 |
| $3_1$ 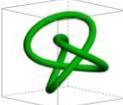 | 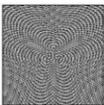 | 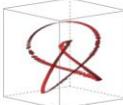 | 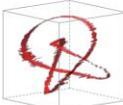 | 3 | $1-t+t^2$ |
| $0_1^2$ 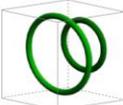 | 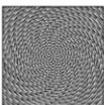 | 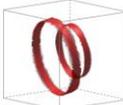 | 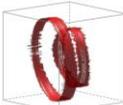 | - | 2 |
| $2_1^2$ 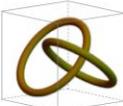 | 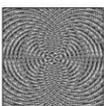 | 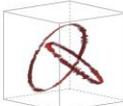 | 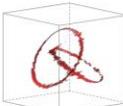 | 1 | 0 |
| $4_1^2$ 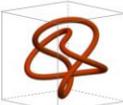 | 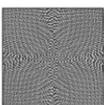 | 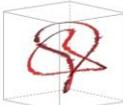 | 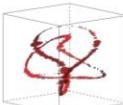 | 2 | $1-t+t^2-t^3$ |
| $0_1^3$ 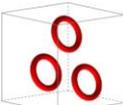 | 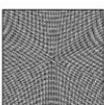 | 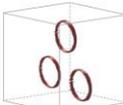 | 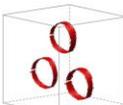 | - | 3 |
| $4_1$ 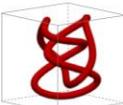 | 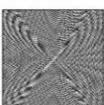 | 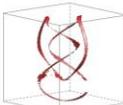 | 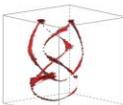 | 2 | $1-3t+t^2$ |
| $5_1$ 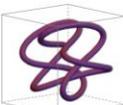 | 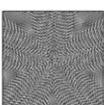 | 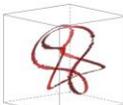 | 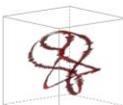 | 5 | $1-t+t^2-t^3+t^4$ |
| $6_1^2$ 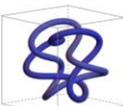 | 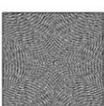 | 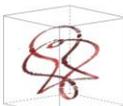 | 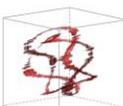 | 3 | $1-t+t^2-t^3+t^4-t^5$ |
| $6_3^3$ 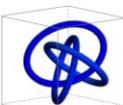 | 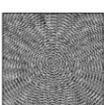 | 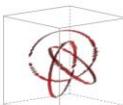 | 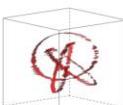 | 2/3 | 0 |

**Figure S4 (part1/2). Generation of optical knots and links.** First columns: the 3D topological structures of knots and links. Second columns: the phase distributions calculated with Eq. (S1). The third columns: the simulation results of optical reconstructed knotted and linked structures. The fourth columns: the experimental results of optical reconstructed knotted and linked structures. The fifth columns: the linking numbers of knots and links. The sixth columns: the Alexander polynomials of knots and links.

| Knot/Link: | arg[$C_A(\varepsilon,\eta)$] PCA | Simulation results | Experimental results | Linking number | Alexander polynomials |
|---|---|---|---|---|---|
| $6_1$ | 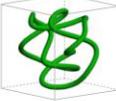 | 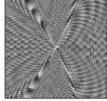 | 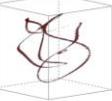 | 4 | $2 - 5t + 2t^2$ |
| $6_1^3$ | 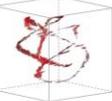 | 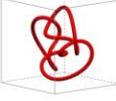 | 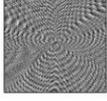 | 2 | $3 - 6t + 3t^2$ |
| $6_2^3$ | 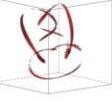 | 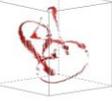 | 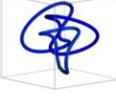 | 0 | $t - 4t^2 - 6t^3 + 4t^4 - t^5$ |
| $7_1$ | 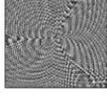 | 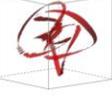 | 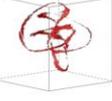 | 7 | $1 - t + t^2 - t^3 + t^4 - t^5 + t^6$ |
| $7_4$ | 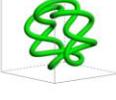 | 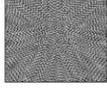 | 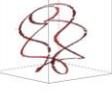 | 7 | $4 - 7t + 4t^3$ |
| $8_1^2$ | 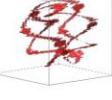 | 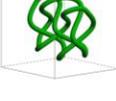 | 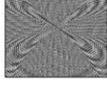 | 4 | $1 - 2t + 3t^2 - 3t^3 + 3t^4 - 3t^5 + 2t^6 - t^7$ |
| $8_{16}$ | 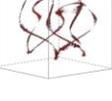 | 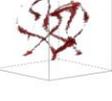 | 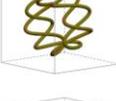 | 2 | $1 - 4t + 6t^2 - 9t^3 + 8t^4 - 4t^5 + t^6$ |
| $8_{18}$ | 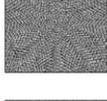 | 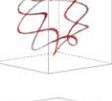 | 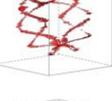 | 0 | $1 - 5t + 10t^2 - t13^3 + 10t^4 - 5t^5 + t^6$ |
| $9_1$ | 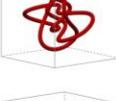 | 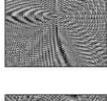 | 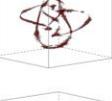 | 9 | $1 - t + t^2 - t^3 + t^4 - t^5 + t^6 - t^7 + t^8$ |
| $0_1^4$ | 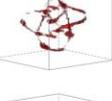 | 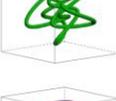 | 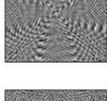 | - | 4 |

**Figure S4 (part2/2). Generation of optical knots and links.** First columns: the 3D topological structures of knots and links. Second columns: the phase distributions calculated with Eq. (S1). The third columns: the simulation results of optical reconstructed knotted and linked structures. The fourth columns: the experimental results of optical reconstructed knotted and linked structures. The fifth columns: the linking numbers of knots and links. The sixth columns: the Alexander polynomials of knots and links.

**Supporting Information Note 3. Topological holography results for the array of $6_1^2$ linked lines**

In the main text, the optical topological holography with $3_1$ knot has been presented. Topological holography with other kinds of knotted/linked structures can also be realized. Here, another example is offered. When $\mathbf{K}(t)$ stands for the $6_1^2$ link, the intensity distributions at different transverse planes in $(x, y, z)$ coordinate are calculated using Eq. (2) in the main text and shown in Figure S5a. The enlarged view of part at $z = 0$ plane (Figure S5b) shows that there are twelve extreme maximum points on this cross section. The simulation results of array of isolated optical $6_1^2$ linked lines are shown in Figure S5c. The corresponding experimental results are shown in Figure S5d-f. It can be seen that our experimental results are consistent with the theoretical ones by comparing the Figure S5c with 5f.

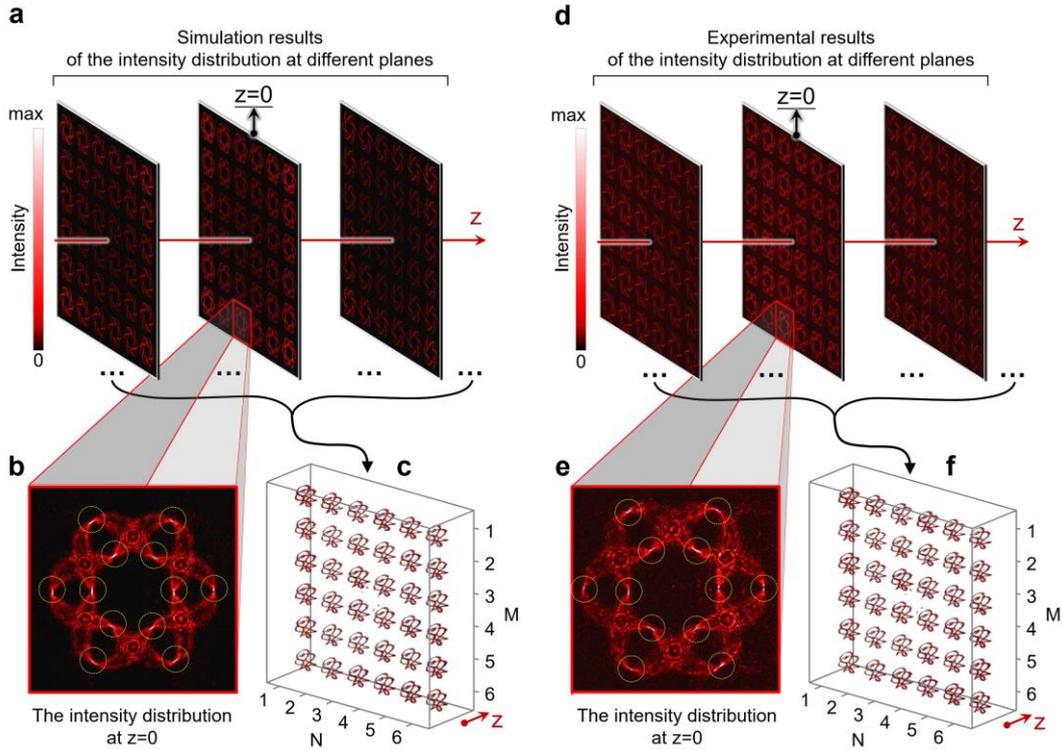

**Figure S5. Simulation and experimental results of optical topological holography. a**, Simulation results of the intensity distributions at different transverse planes in $(x, y, z)$ coordinate. **b**, The enlarged view of one part at $z = 0$ plane shown in **a**. **c**, Simulation results of optical topological holography. **d-f**, Corresponding experimental results of **a-c**, respectively.

**Supporting Information Note 4. Robustness of topological structure to the noise from random phase and deformation of hologram.**

Robustness of generation of optical knots and links to the amplitude distribution of incident field has been demonstrated (Supplementary Note 2). In the following, we discuss the effect of random phase noise and deformation of hologram. Here, we take the knot $3_1$ as an example. We add a random phase within [0, n] to the phase distribution $\arg[C_A(\varepsilon, \eta)]$, as shown in the first row of the Figure S6 with n = 0, 5, 10, 15, and 20. The light field that contains the 3D

$3_1$ knotted topological structure are calculated. The optical topological knotted structures are reconstructed under conditions of different random phase and shown in the second row in the Figure S6. The third row of the Figure S6 shows the corresponding intensity distribution at plane of z = 0. These results show that the generation of optical knocked and links has good resistance to random phase. Although the noise of the reconstructed optical topological knotted structures increases along with the range of random phase, the $3_1$ knot can still be distinguished when the transformation range of random phase increases to [0,15].

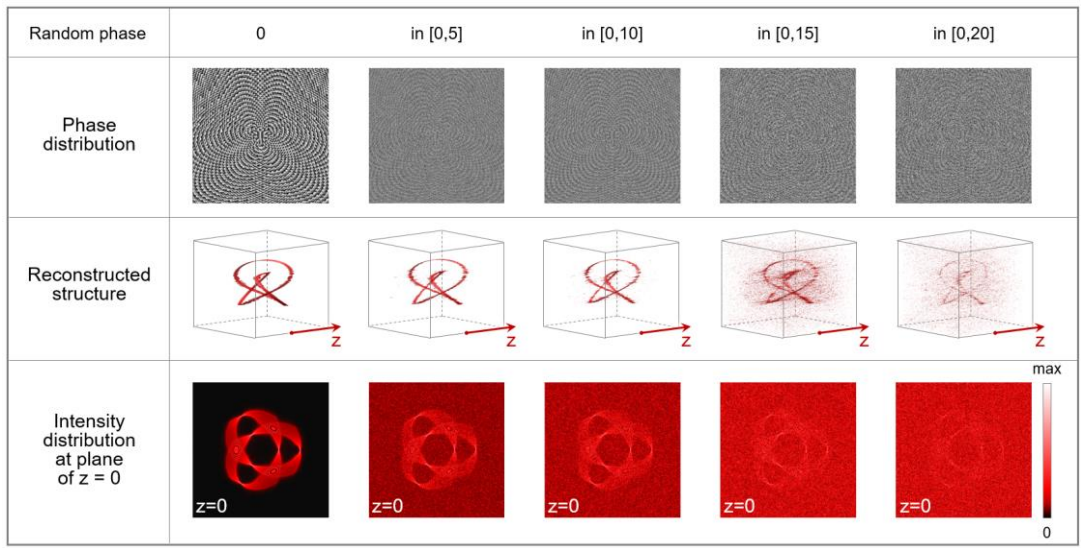

**Figure S6. Effect of random phase on the generation of optical knots and links.** First row: phase distribution with random phase within [0, n] with n = 0, 5, 10, 15, and 20. Second and third rows represent the reconstructed optical topological knotted structures and the intensity distribution at plane of z = 0 for conditions of different random phase.

Next, I studied the influence of hologram deformation on the generated topological structures by changing the aspect ratio of the hologram. For the purposes of comparison, we set the aspect ratio of phase distribution to be 10:10, 10:9, 10:8, 10:7, or 10:6 as shown in the first row of the Figure S7. The corresponding topological knotted structures are reconstructed and shown in the second row in the Figure S7. The third row of the Figure S7 shows the corresponding intensity distribution at plane of z = 0. These results show that hologram deformation changes the shape of the 3D optical knotted structure, but does not change the topological invariants (the linking number and Alexander polynomial) of knot, which mean that the type of reconstructed knot keeps unchanged.

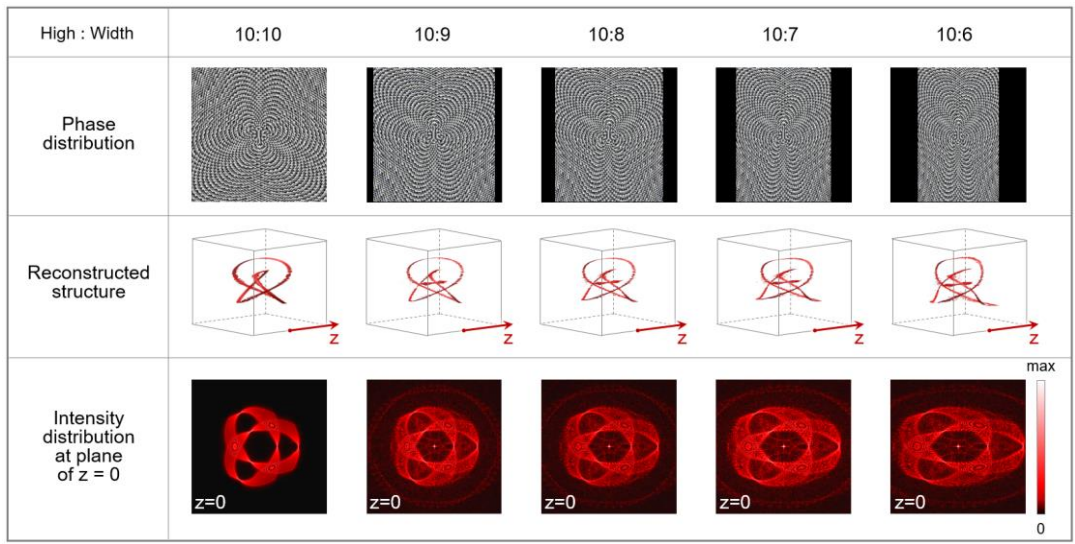

**Figure S7. Effect of hologram deformation on the generation of optical knots and links.** First row: The aspect ratio of phase distribution is set to be 10:10, 10:9, 10:8, 10:7, or 10:6. Second and third rows represent the reconstructed optical topological knotted structures and the intensity distribution at plane of z = 0 for conditions of different random phase.